\documentstyle[12pt]{article}

\begin{document}

\baselineskip = 20pt

\title{\bf Exact Effective Action for ($1+1$ Dimensional) Fermions in
  an Abelian Background at Finite Temperature}
\medskip

\author{Ashok Das\\
Department of Physics and Astronomy \\
University of Rochester\\
Rochester, NY 14627, USA\\
\\
and\\
A. J. da Silva\\
Instituto de F\'\i sica\\
Universidade de S\~ao Paulo\\
Caixa Postal 66318\\
S\~ao Paulo, SP 05315-970, BRASIL\\
\\}
\date{}
\maketitle
\medskip

\begin{abstract}

In an effort to further understand the structure of effective actions
for fermions in an external gauge background at finite temperature, we
study the example of $1+1$ dimensional fermions interacting with an
arbitrary Abelian gauge field. We evaluate the effective action
exactly at finite temperature. This effective action is non-analytic as
is expected at finite temperature. However, contrary to the structure
at zero temperature and contrary to naive expectations, the effective
action at finite temperature has interactions to all (even)
orders (which, however, do not lead to any quantum corrections). The
covariant  structure thus obtained may prove useful in
studying $2+1$ dimensional models in arbitrary backgrounds. We also
comment briefly on the solubility of various $1+1$ dimensional models
at finite temperature.

\end{abstract}

\vfill

\pagebreak

\section{Introduction:}

\par

Finite temperature introduces various new features \cite{das1} into
quantum field 
theories that we are not used to at zero temperature. Thus, for
example, it is known that various amplitudes as well as the effective
actions can become non-analytic at finite temperature \cite{{das1},
{weld1}, {das2}} (beyond $0+1$
dimensions)  which is
connected with the existence of additional channels of reactions
possible in a thermal medium. There are also various subtleties that
arise, such as the modified
Feynman combination  formula \cite{{das1}, {weld2}, {das3}}, because the
propagators do not have simple analytic behavior at finite
temperature. More recently, it is also found that the effective action
 at finite temperature can be non-extensive \cite{dunne} unlike at zero
temperature. Thus, for the $0+1$ dimensional fermions interacting with
an Abelian gauge field, the effective action at finite temperature
becomes a non-polynomial function of $(\int dt\, A)$ where $A$ represents
the external, Abelian gauge field. This new structure of the effective
action has led to a successful understanding of the question of large
gauge invariance, at finite temperature, in this model. This model has
properties similar to that of the $2+1$ dimensional fermions
interacting with an arbitrary external gauge field in the sense that
the radiative corrections induce a Chern-Simons term whose coefficient
is a continuous function of temperature \cite{{das4}, {aitch1}}
and, therefore,  incompatible
with the quantization condition necessary for large gauge invariance
to hold \cite{deser1}. The study of the $0+1$ dimensional model
suggests a way  for
the understanding of the question of large gauge invariance in the
$2+1$ dimensional model at finite temperature and there have been
several attempts to generalize the results of the $0+1$ dimensional
model to the case of the $2+1$ dimensional model \cite{{deser2},
{fosco}, {aitch2}}. However, these
attempts, typically, deal with very specific gauge backgrounds and a
systematic study of the effective action for the $2+1$ dimensional
fermions interacting with an arbitrary gauge background is still
lacking.

While  various properties of the $0+1$ dimensional model, at finite
temperature, are quite well understood \cite{{das5}, {das6}}, it
is not at all obvious how
the structure should be generalized to higher dimensions. For one thing,
in $0+1$ dimensions, there is only one component for the gauge field
and, consequently, it is not clear what would be the appropriate
covariant structure that would generalize to higher
dimensions. Second, as pointed out earlier, at finite temperature,
the effective action can become non-analytic beyond $0+1$ dimensions
and this makes any generalization of the results of the $0+1$
dimensional model (where there is no problem of non-analyticity) to
higher dimensions additionally tricky. For these reasons, we have
chosen to study, in this paper, a model of intermediate complexity,
namely, the $1+1$ dimensional fermions interacting with an arbitrary
external Abelian gauge field with the hope that it would shed light on
some of the issues raised.

We consider massless fermions interacting with an external Abelian
gauge field which, of course, can be exactly solved at zero
temperature (leads to only quadratic terms in the effective action)
and is associated  with the solubility of various two
dimensional models \cite{{hagen1}, {bass}, {das7},
{hagen2}}. This model, of course, is not directly related
to the question of large gauge invariance, but it is the structure
of the effective action at finite temperature that we are interested
in. It is well known that the chiral anomaly (which is responsible for
the solubility of the model) of this model is not changed in the
presence of temperature \cite{{das1}, {jackiw}, {das8}}. In a gauge
invariant regularization  (which is what we
will use, but let us emphasize that the finite temperature
calculations are all finite and it is the zero temperature calculation
that needs a regularization), therefore, it would seem, {\it a priori}, that there would be no
temperature dependent corrections to the effective action. Namely, if
there is a temperature dependent correction, $\Gamma^{(\beta)}$, it must
satisfy 

\begin{equation}
\partial_{\mu}{\delta \Gamma^{(\beta)}\over \delta A_{\mu}} = 0 =
\epsilon_{\mu\nu}\partial^{\mu}{\delta \Gamma^{(\beta)}\over \delta
  A_{\nu}}\label{a_1} 
\end{equation}
With the usual assumptions of locality, then, it would follow that
$\Gamma^{(\beta)}=0$. However, as we have learnt from the study of the $0+1$
dimensional model \cite{dunne}, the effective action at finite temperature can be
non-extensive in which case, it is not necessary for $\Gamma^{(\beta)}$ to
vanish. In fact, taking from the results of the $0+1$ dimensional
model \cite{{dunne}, {das5}}, we note that a simple, non-extensive
quadratic  term in the
effective action of the form ($c$ is a constant)

\begin{equation}
\Gamma_{q} = c \left(\int d^{2}x\,A_{\mu}(x)\right)\left(\int
  d^{2}y\,A^{\mu}(y)\right)\label{a_2} 
\end{equation}
can give rise to a current that has vanishing divergence and curl. Thus, we
would like to systematically study the structure of the effective
action for this $1+1$ dimensional model at finite temperature.

The paper is organized as follows. In section {\bf 2}, we recapitulate
briefly the structure of the quadratic effective action at zero
temperature. We, then, generalize a  theorem of zero temperature
\cite{das9} which shows that the 
effective action would continue to be quadratic even at finite
temperature. In section {\bf 3}, we evaluate the two point function to
show  that the quadratic term does have a temperature dependent
correction which does not alter the current conservation and the
anomaly of the theory. The non-analytic structure of this correction is
pointed out and the quadratic temperature dependent term of the
effective action is expressed in a manifestly covariant fashion. As we
have mentioned earlier, generally, various subtleties arise at finite
temperature. Consequently, in section {\bf 4}, we calculate explicitly
the 3-point and the 4-point functions. While the 3-point function
vanishes (in fact, all the odd point functions must vanish because of
charge conjugation invariance), surprisingly, the 4-point function is
nontrivial at finite temperature and correspondingly, there is a
quartic term in the temperature dependent effective action. We
clarify the reason for the 
failure of the general theorem. We also obtain the general form for
the $2n$-point function which is nontrivial and, thereby, determine
the complete effective action at finite temperature. We show that, in
a dynamical gauge theory, these additional interactions, however, do
not generate any quantum mechanical correction which is yet a new
feature at finite temperature. In section {\bf  5}, we make some brief
comments  about the solubility of various two
dimensional models at finite temperature and present some brief
conclusions in section {\bf 6}.

\section{General Theorem:}

In this section, we will recapitulate briefly the structure of the
effective action at zero temperature and prove a general theorem
following the method at zero temperature \cite{das9} concerning the
structure  of
the effective action at finite temperature. Let us note that we are
interested in the $1+1$ dimensional model described by the Lagrangian
density 
 
\begin{equation}
{\cal L} = {\overline {\psi}}\gamma^{\mu}(i\partial_{\mu} - e
A_{\mu})\psi.\label{a_3} 
\end{equation}
We use $\eta^{\mu\nu}=(+,-)$ with $\mu,\nu=0,1$. Although not
necessary, a representation for the Dirac matrices can be chosen to be
$\gamma^{0}= \sigma_{2}$, $\gamma^{1}= i\sigma_{1}$. In $1+1$
dimensions, the gamma matrices further satisfy the identity
\begin{equation}
\gamma^{\mu}\gamma^{\nu} = \eta^{\mu\nu} + \epsilon^{\mu\nu}
\gamma_{5}\label{a_4} 
\end{equation}
where $\gamma_{5}=\gamma^{0}\gamma^{1}$ and $\epsilon^{\mu\nu}$ is the
anti-symmetric Levi-Civita tensor with $\epsilon^{01}=1$. Integrating
out the fermions in the path integral leads to the effective action

\begin{eqnarray}
\Gamma[A]&=&-i\ln{\det\left( i\partial\!\!\!\slash-eA\!\!\!\!\slash\right)
\over \det(i\partial\!\!\!\slash)}\nonumber\\
&=&-i\ln\det\left( 1-eS(p)A\!\!\!\!\slash\right)\nonumber\\
&=&-i{\rm Tr}\ln\left( 1-eS(p)A\!\!\!\!\slash\right).\label{a_5}
\end{eqnarray}
Here, we have normalized the effective action so that it vanishes for
vanishing external field. The \lq\lq Tr''in (\ref{a_5}) stands for the
trace in a complete basis as well as a Dirac trace and $S(p)$ is the
propagator for the fermion.

We can expand the logarithm in (\ref{a_5}) which leads to a power series
representation for the effective action
\begin{equation}
\Gamma[A] = i{\rm Tr}\left((eS(p)A\!\!\!\!\slash) +
  {(e^{2}S(p)A\!\!\!\!\slash S(p)A\!\!\!\!\slash)\over 2} + \cdots
  \right)\label{a_6}
\end{equation}
Here $S(p)$ and $A(x)$ are supposed to be non-commuting operators and
the effective action, in general, contains an infinite number of
terms. However, let us note that, in $1+1$ dimensions, we can
decompose the vector field as
\begin{equation}
A_{\mu} = {1\over e}(\partial_{\mu}\sigma +
\epsilon_{\mu\nu}\partial^{\nu}\phi)\label{a_7} 
\end{equation}
and that at zero temperature, the fermion propagator has the form
($i\epsilon$ prescription is understood)
\begin{equation}
S(p) = {1\over p\!\!\!\slash}.\label{a_8}
\end{equation} 
It is, therefore, clear that at zero temperature, we can write (using
(\ref{a_4})) 
\begin{eqnarray}
eS(p)A\!\!\!\!\slash & = & {1\over p\!\!\!\slash}(\partial\!\!\!\slash\sigma -
\gamma_{5}\partial\!\!\!\slash\phi)\nonumber\\
 & = & {1\over p\!\!\!\slash}(-i)[p\!\!\!\slash , \sigma ] +
 \gamma_{5}{1\over p\!\!\!\slash} (-i)[p\!\!\!\slash , \phi
 ]\nonumber\\
 & = & -i\sigma +i{1\over p\!\!\!\slash}\sigma p\!\!\!\slash -
 i\gamma_{5}\phi + i\gamma_{5}{1\over p\!\!\!\slash}\phi p\!\!\!\slash
 .\label{a_9}
\end{eqnarray}
It now follows from this that
\begin{equation}
(eS(p)A\!\!\!\!\slash)^{2} = -i\left[\sigma + \gamma_{5}\phi ,
  (eS(p)A\!\!\!\!\slash)\right]\label{a_10}
\end{equation}
which gives
\begin{equation}
{\rm Tr}(eS(p)A\!\!\!\!\slash)^{n+1} = -{i\over n}{\rm Tr}\left[\sigma
  + \gamma_{5}\phi , (eS(p)A\!\!\!\!\slash)^{n}\right]\label{a_11}
\end{equation}
For $n>1$, these integrals are convergent and hence one can use the
cyclicity of the trace to conclude that all the terms in the effective
action in (\ref{a_6}) which are higher order than the quadratic vanish at zero
temperature. Even the linear term in (\ref{a_6}) vanishes because of
the odd nature of the integrand. The quadratic term in the effective
action can be evaluated in a straightforward manner and a gauge
invariant regularization gives the complete effective action at
zero temperature to be (although there is a one parameter
freedom of regularization, we will use a gauge invariant
regularization for simplicity)
\begin{equation}
\Gamma^{(0)}[A] = {e^{2}\over 2\pi}\int
d^{2}x\,A_{\mu}\left(\eta^{\mu\nu}-{\partial^{\mu}\partial^{\nu}\over
    \partial^{2}}\right)A_{\nu} .\label{a_12}
\end{equation}

The above proof can also be easily generalized to finite
temperature. Let us recall that, at finite temperature, the fermion
propagator has the form \cite{das1}
\begin{equation}
S(p) = {1\over p\!\!\!\slash} + 2i\pi p\!\!\!\slash n(|p^{0}|) \delta
(p^{2})\label{a_13} 
\end{equation}
where $n(|p^{0}|)$ represents the fermion distribution function
($\beta = {1\over kT}$, $k = $ Boltzmann constant)
\begin{equation}
n(|p^{0}|) = {1\over e^{\beta |p^{0}|} + 1}\label{a_14}
\end{equation}
We note that, with the parameterization in (\ref{a_7}), we can write
\begin{eqnarray}
eS(p)A\!\!\!\!\slash & = & -i S(p)\{[p\!\!\!\slash , \sigma] +
\gamma_{5}[p\!\!\!\slash ,\phi ]\}\nonumber\\
 & = & -i\left[\sigma - {1\over p\!\!\!\slash}\sigma p\!\!\!\slash
   -2i\pi p\!\!\!\slash n(|p^{0}|)\delta (p^{2})\sigma
   p\!\!\!\slash\right.\nonumber\\ 
 &   &\;\;\;+\left.\gamma_{5}\left\{\phi - {1\over p\!\!\!\slash}\phi
     p\!\!\!\slash - 2i\pi p\!\!\!\slash n(|p^{0}|)\delta (p^{2})\phi
     p\!\!\!\slash \right\}\right]\label{a_15}
\end{eqnarray}
Although this is very different from the structure at zero
temperature, namely, eq. (\ref{a_9}), it still leads to
\begin{equation}
(eS(p)A\!\!\!\!\slash)^{2} = -i\left[\sigma + \gamma_{5}\phi ,
  (eS(p)A\!\!\!\!\slash) \,\right]\label{a_16}
\end{equation}
Consequently, as in the zero temperature case, we can write
\begin{equation}
{\rm Tr}(eS(p)A\!\!\!\!\slash)^{n+1} = -{i\over n}{\rm Tr}\left[\sigma +
  \gamma_{5}\phi , (eS(p)A\!\!\!\!\slash)^{n}\right]\label{a_17} 
\end{equation}
which, therefore, would seem to suggest that much like the zero
temperature case, the finite temperature effective action is at most
quadratic in the external fields. In fact, the linear terms vanish by
anti-symmetry of the integrand and so, it would seem that the only
modification that temperature might induce is at most to change the
two point function.

\section{ Two Point Function:}

The calculation of the two point function is not really very
difficult. There is only one Feynman diagram to evaluate which has the
form ($i\epsilon$ prescription is understood)
\begin{eqnarray}
i\Gamma^{\mu\nu}(p^{0}, p^{1}) & = & -e^{2} \int {d^{2}k\over
  (2\pi)^{2}}\,\{k_{+}^{\mu}(k+p)_{+}^{\nu} +
  k_{-}^{\mu}(k+p)_{-}^{\nu}\}\nonumber\\
 &  & \times\left({1\over k^{2}} + 2i\pi n(|k^{0}|)\delta
  (k^{2})\right)\!\!\!\left({1\over (k+p)^{2}} + 2i\pi n(|k^{0}+p^{0}|)\delta
  ((k+p)^{2})\right)\label{a_18}
\end{eqnarray}
where we have defined
\begin{equation}
k_{\pm}^{\mu} = (\eta^{\mu\nu} \pm
\epsilon^{\mu\nu})k_{\nu}\label{a_19}
\end{equation} 
The zero temperature part, of course, can be read out from
eq. (\ref{a_12}) and, therefore, we would concern ourselves, in this
section, only with possible temperature dependent corrections to the two
point function. We note from the definition in eq. (\ref{a_19}) that
only two independent tensor structures arise from (\ref{a_18}). The
evaluation is straightforward and we have
\begin{eqnarray}
i\Gamma^{00(\beta)}(p^{0}, p^{1}) & = &
i\Gamma^{11(\beta)}(p^{0}, p^{1})\nonumber\\
 & = & \left(\delta (p_{-}) + \delta (p_{+})\right) I_{2}\nonumber\\
 &  &  \nonumber\\
i\Gamma^{01(\beta)}(p^{0}, p^{1}) & = &
i\Gamma^{10(\beta)}(p^{0}, p^{1})\nonumber\\
 & = & \left(\delta(p_{-}) - \delta (p_{+})\right) I_{2}\label{a_20}
\end{eqnarray}
Here, we have defined
\begin{equation}
p_{\pm} = p^{0} \pm p^{1}\label{a_21}
\end{equation}
and
\begin{eqnarray}
I_{2} & = & {(2ie\pi)^{2}\over 2} \int
{dk^{1}\over
  (2\pi)^{2}}\,\left[\epsilon(k^{1})\epsilon(k^{1}+p^{1})\{n(|k^{1}|) +
  n(|k^{1}+p^{1}|)\right.\nonumber\\
  &   & \hspace{1.5in}\left. - 2 n(|k^{1}|)n(|k^{1}+p^{1}|)\}\right]\label{a_22}
\end{eqnarray}
Here $\epsilon(x)$ stands for the alternating step function.

Thus, we see that there is indeed a temperature dependent correction
to the two point function. Furthermore, there are several things to
note from the structure of the temperature dependent part in
(\ref{a_20}). First, it is easy to verify from (\ref{a_20}) that
\begin{equation}
p_{\mu} \Gamma^{\mu\nu (\beta)} (p^{0}, p^{1}) = 0 = p_{\nu}
\Gamma^{\mu\nu (\beta)}(p^{0}, p^{1})\label{a_23}
\end{equation}
so that this additional correction is transverse as gauge invariance
would require. Furthermore, it is also equally straightforward to
check that
\begin{equation}
\epsilon_{\mu\nu} p^{\mu}\Gamma^{\nu\lambda (\beta)}(p^{0}, p^{1})
= 0 = \epsilon_{\mu\lambda}p^{\mu}\Gamma^{\nu\lambda (\beta)}(p^{0}, p^{1})
\label{a_24}
\end{equation}
In other words, this temperature dependent correction would lead to a
modification in the current which has vanishing divergence as well as
curl (and yet is not trivial).

The two point function, in eq. (\ref{a_20}) is clearly non-analytic at
the origin in the $(p^{0}, p^{1})$ plane which is best seen by writing
\[
p^{0} = \alpha p^{1}
\]
and noting that in the limit $p^{1}\rightarrow 0$, the amplitude
depends on the parameter $\alpha$. This is the well known
non-analyticity in the two point function that is expected at finite
temperature. However, it is interesting to note that the
non-analyticity, in this case, manifests essentially  in the
structure of the delta functions which is also quite crucial for the
current conservation as well as the vanishing of the anomaly (which is
clear from eqs. (\ref{a_23}, \ref{a_24})). We suspect that this is a
structure that may  generalize to higher dimensions in a calculation
with an arbitrary gauge background. (We would like to point out here that this dependence on the delta function is a particular generalization of the $0+1$ dimensional result \cite{{das5}, {das6}} where the two point function has only one component and is proportional to $\delta(p)$.)

The two point function, of course, can be expressed in  a more
covariant form. The standard way to do this is to introduce a velocity
for the heat bath, $u^{\mu}$, such that \cite{weldon3}
\[
u^{\mu}u_{\mu} = 1
\]
Without going into too much detail, let us note that every four-vector
can now be decomposed along parallel and perpendicular directions to
$u^{\mu}$ as \cite{das10}
\begin{eqnarray}
k^{\mu} & = & \Omega u^{\mu} - \epsilon^{\mu\nu}u_{\nu}\overline{k}\nonumber\\
p^{\mu} & = & \omega u^{\mu} - \epsilon^{\mu\nu}u_{\nu}\overline{p}\label{a_25}
\end{eqnarray}
where the Lorentz invariant quantities $\Omega$, $\omega$,
$\overline{k}$ and $\overline{p}$
are defined by
\begin{eqnarray}
\Omega = k^{\mu}u_{\mu} & \; ; \;  & \overline{k} =
\epsilon^{\mu\nu}k_{\mu}u_{\nu}\nonumber\\
\omega = p^{\mu}u_{\mu} & \; ; \;  & \overline{p} =
\epsilon^{\mu\nu}p_{\mu}u_{\nu}\label{a_26}
\end{eqnarray}
We can also define the component of the velocity four-vector
perpendicular to $p^{\mu}$ as
\begin{equation}
\overline{u}^{\mu}(p) = u^{\mu} - {\omega\over
  \overline{p}}\epsilon^{\mu\nu}u_{\nu} \label{a_27}
\end{equation}

The calculation of the two point function can be easily carried out in
terms of these variables and the temperature dependent correction has
the form
\begin{equation}
i\Gamma^{\mu\nu (\beta)}(\omega, \overline{p}) = \left(\delta (\omega -
  \overline{p}) + \delta (\omega +
  \overline{p})\right)\overline{u}^{\mu}(p)\overline{u}^{\nu}(-p)\overline{I}_{2}\label{a_28} 
\end{equation}
where
\begin{equation}
\overline{I}_{2} = {(2ie\pi)^{2}\over 2}\int {d\overline{k}\over
  (2\pi)^{2}}\,\epsilon(\overline{k})\epsilon(\overline{k}+\overline{p})\left[n(|\overline{k}|) + n(|\overline{k}+\overline{p}|) - 2 n(|\overline{k}|)n(|\overline{k}+\overline{p}|)\right]\label{a_29}
\end{equation}
This single tensor structure indeed generates the two independent
structures noted in (\ref{a_20}) and, in fact, reduces to them in the
rest frame of  the heat bath for which $u^{\mu}
= (1, 0)$. We also note here that the transversality of the two point
function follows trivially from the fact that $\overline{u}^{\mu}(p)$
is transverse to $p^{\mu}$. The vanishing of the curl, however, does
not follow from the transversality of $\overline{u}^{\mu}$; rather, it
is a consequence of the delta function structure of the two point
function. (Note also that, in addition to the delta functions being
non-analytic, $\overline{u}^{\mu}$ also depends on the direction along
which we approach the origin.)

Once we have a covariant expression for the temperature dependent
correction to the two point function, we can easily write the
additional quadratic term that would be generated at finite
temperature in the effective action,
\begin{eqnarray}
\Gamma_{2}^{(\beta)} & = & {1\over 2!}\int {d\omega d\overline{p}\over
  (2\pi)^{2}}\,A_{\mu}(p) (i\Gamma^{\mu\nu (\beta)})
  A_{\nu}(-p)\nonumber\\
& = &  {1\over 2!}\int {d\omega d\overline{p}\over
  (2\pi)^{2}}\,(\overline{u}\cdot A)(p) (\overline{u}\cdot
  A)(-p) \overline{I}_{2}\left(\delta (\omega - \overline{p}) + \delta
  (\omega + \overline{p})\right)\label{a_30}
\end{eqnarray}
As is obvious, this action is highly nonlocal although it does not
have the non-extensive structure found in the $0+1$ dimensional
model. It is not obvious to us, at this point,  whether the
non-extensive structure  is a
special feature in odd space-time dimensions or simply an accidental
feature of the $0+1$ dimensional model. We would also like to
comment here that the structure of the quadratic term in (\ref{a_30})
is manifestly gauge invariant because of the transversality of
$\overline{u}^{\mu}$. While it is not obvious, it can be easily checked that
the same structure is also invariant for non-Abelian gauge fields
which may be an interesting thing to note for generalizations to
non-Abelian theories in higher dimensions.

\section{Higher Point Functions:}

The general proof of section {\bf 2} would seem to suggest that the
quadratic term is all the correction that temperature would induce in
the effective action. However, as we have pointed out earlier, there
are often subtleties that arise at finite temperature. As a result, it
is always useful to check explicitly for such possibilities. In what
follows, we would, therefore, like to check explicitly if the 3-point
and the 4-point functions, for example, continue to vanish at finite
temperature.

The calculation of the 3-point function is only slightly more
difficult than the two point function. In this case, the
amplitude involves evaluating two Feynman diagrams. If we denote the
two independent external momenta by $p$ and $q$, then, the two
independent diagrams would correspond to exchanging $p\leftrightarrow
q$ (of course, with the appropriate interchange of the tensor
indices). The 3-point function has the structure
\begin{eqnarray}
i\Gamma^{\mu\nu\lambda}(p, q) & = & -e^{3} \int \left. {d^{2}k\over
    (2\pi)^{2}}\,
    \right[\left(k_{+}^{\mu}(k+p)_{+}^{\nu}(k+p+q)_{+}^{\lambda} +
    k_{-}^{\mu}(k+p)_{-}^{\nu}(k+p+q)_{-}^{\lambda}\right)\nonumber\\
  &   & \times\left({1\over k^{2}}+ 2i\pi n(|k^{0}|)\delta
    (k^{2})\right)\left({1\over (k+p)^{2}} + 2i\pi
    n(|k^{0}+p^{0}|)\delta ((k+p)^{2})\right)\label{a_31}\\
  &   & \left. \times\left({1\over (k+p+q)^{2}} + 2i\pi
    n(|k^{0}+p^{0}+q^{0}|)\delta ((k+p+q)^{2})\right) +
    (p,\nu\leftrightarrow q,\lambda)\right]\nonumber
\end{eqnarray}
where we have used the definitions in (\ref{a_19}). Once again, we can
easily see that the temperature dependent corrections have only two
independent structures, the ones with an even number of space-like
indices (they are equal) and the ones with an odd number of space-like
indices (which are also equal). The two independent structures can be
evaluated to have the simple forms
\begin{eqnarray}
i\Gamma^{000 (\beta)} (p, q) & = & \left(\delta (p_{-})\delta (q_{-})
  + \delta (p_{+})\delta (q_{+})\right) I_{3}\nonumber\\
i\Gamma^{001 (\beta)} (p, q) & = & \left(\delta (p_{-})\delta (q_{-})
  - \delta (p_{+})\delta (q_{+})\right) I_{3}\label{a_32}
\end{eqnarray}
where
\begin{eqnarray}
I_{3}(p, q) & = & {(2ie\pi)^{3}\over 2^{2}}\int  {dk^{1}\over
    (2\pi)^{2}}\,
    \left[\epsilon(k^{1})\epsilon(k^{1}+p^{1})\epsilon(k^{1}+p^{1}+q^{1})\right.\nonumber\\
  &   & \times\left\{-\left(n(|k^{1}|) + n(|k^{1}+p^{1}|) +
    n(|k^{1}+p^{1}+q^{1}|)\right)\right.\nonumber\\
  &   & \;\;\;\; + 2\left(n(|k^{1}|)n(|k^{1}+p^{1}|) +
    n(|k^{1}|)n(|k^{1}+p^{1}+q^{1}|)\right.\nonumber\\
  &   & \;\;\;\;\;\;\;\;\;\;\;\;\left.  +
    n(|k^{1}+p^{1}|)n(|k^{1}+p^{1}+q^{1}|)\right) \nonumber\\
  &   &
    \left.\left. \;\;\;\; -4n(|k^{1}|)n(|k^{1}+p^{1}|)n(|k^{1}+p^{1}+q^{1}|)\right\} + (p\leftrightarrow q)\right]\label{a_33}
\end{eqnarray}
With an appropriate change of variables, it is easy to see that
$I_{3}$ vanishes because of the anti-symmetry of an odd number of
alternating  step
functions. (Namely, the two diagrams exactly cancel each other.) In
fact, one can show,  in general, that all the odd-point
functions vanish because of charge conjugation invariance in the
theory. Namely, the
Lagrangian density in (\ref{a_3}) is invariant under
\[
\psi\rightarrow \eta\, C\,\overline{\psi}^{T}; \;\;\;\; A_{\mu}\rightarrow
-A_{\mu}
\]
where $\eta$ is a phase and $C$ represents the charge conjugation
matrix. This invariance requires that the effective action can only
depend on an even number of $A_{\mu}$ fields. However, in spite of this
general result, we went through the explicit calculation to show the
generalization of the structure of the two point function to the case
of the three point function. (Had the three point function not
vanished, transversality as well as vanishing anomaly would have
required the structure to be a generalization of the two point
function in the form in (\ref{a_32}).)

Let us next turn to the 4-point function. This is much more involved than
the 3-point function. Furthermore, there are now six diagrams to
be evaluated. However, each of them has the generic form ($p$, $q$ and $r$ are
the independent external momenta)
\begin{eqnarray*}
  & = & -e^{4}\int \left. {d^{2}k\over
  (2\pi)^{2}}\,\right[\left\{k_{+}^{\mu}(k+p)_{+}^{\nu}(k+p+q)_{+}^{\lambda}(k+p+q+r)_{+}^{\rho}\right.\\  
  &   & \hspace{1in}\left. + k_{-}^{\mu}(k+p)_{-}^{\nu}(k+p+q)_{-}^{\lambda}(k+p+q+r)_{-}^{\rho}\right\}\\
  &   & \times\left({1\over k^{2}} + 2i\pi n(|k^{0}|)\delta
  (k^{2})\right)\left({1\over (k+p)^{2}} + 2i\pi
  n(|k^{0}+p^{0}|)\delta ((k+p)^{2})\right)\\
  &   & \times\left({1\over (k+p+q)^{2}} + 2i\pi
  n(|k^{0}+p^{0}+q^{0}|)\delta ((k+p+q)^{2})\right)\\
  &   & \times \left.\left({1\over
  (k+p+q+r)^{2}} + 2i\pi n(|k^{0}+p^{0}+q^{0}+r^{0}|)\delta
  ((k+p+q+r)^{2})\right)\right]
\end{eqnarray*}
The calculation for the temperature dependent part is exactly similar
to the two and the three point functions, but much more
tedious. Adding all the six diagrams,
we find, again, that there are only two independent structures that
arise. There are the ones with an even number of space-like indices
(and they are all equal) and the other kind is for the ones with an
odd number of space-like indices (which are again all equal) with the
forms given by (with appropriate definitions given in (\ref{a_21}))
\begin{eqnarray}
i\Gamma^{0000 (\beta)} (p, q, r) & = & \left(\delta (p_{-})\delta
  (q_{-})\delta (r_{-}) + \delta (p_{+})\delta (q_{+})\delta
  (r_{+})\right) I_{4}\nonumber\\
i\Gamma^{0001 (\beta)} (p, q, r) & = & \left(\delta (p_{-})\delta
  (q_{-})\delta (r_{-}) - \delta (p_{+})\delta (q_{+})\delta
  (r_{+})\right) I_{4}\label{a_34}
\end{eqnarray}
where
\begin{eqnarray}
I_{4} (p, q, r) & = & {(2ie\pi)^{4}\over 2^{3}}\int {dk^{1}\over
  (2\pi)^{2}}\,\left[\epsilon(k^{1})\epsilon(k^{1}+p^{1})\epsilon(k^{1}+p^{1}+q^{1})\epsilon(k^{1}+p^{1}+q^{1}+r^{1})\right.\nonumber\\
  &  & \times\;\;\;\;\left\{n(|k^{1}|) + n(|k^{1}+p^{1}|) +
  n(k^{1}+p^{1}+q^{1}|) +
  n(|k^{1}+p^{1}+q^{1}+r^{1}|)\right.\nonumber\\
  &  & \;\;\;\; - 2 \left(n(|k^{1}|)n(|k^{1}+p^{1}|) + {\rm all\; quadratic\;
  permutations}\right)\nonumber\\
  &  & \;\;\;\; + 4
  \left(n(|k^{1}|)n(|k^{1}+p^{1}|)n(|k^{1}+p^{1}+q^{1}|) + {\rm 
  all\; cubic\; permutations}\right)\nonumber\\
  &  & \left. \;\;\;\; - 8
  n(|k^{1}|)n(|k^{1}+p^{1}|)n(|k^{1}+p^{1}+q^{1}|)n(|k^{1}+p^{1}+q^{1}+r^{1}|)\right\} \nonumber\\
  &  & \left. \;\;\;\; +\; {\rm all\; permutations\; of\;} (p, q,
  r)\right]\label{a_35}
\end{eqnarray}

There are several things to note here. First of all, the 4-point
function involves an even number of alternating step functions and,
therefore, does not vanish unlike the three point function. However,
from the structure in (\ref{a_34}), it is clear that it is divergence
free and does not contribute to the anomaly either. In fact, the
structure is a generalization of the two point and the three point
functions in eqs. (\ref{a_20}) and (\ref{a_32}) respectively. The
non-analyticity continues to be present in the structure of the 4-point
function. Furthermore, we can also write the 4-point function in a
manifestly covariant form as in the case of the two point
function. Let us identify
\begin{equation}
p^{\mu} = p_{1}^{\mu};\;\; q^{\mu} = p_{2}^{\mu};\;\; r^{\mu} =
p_{3}^{\mu}\label{a_36}
\end{equation}
and define, as in eqs. (\ref{a_25}, \ref{a_26}),
\begin{equation}
p_{i}^{\mu} = \omega_{i}u^{\mu} -
\epsilon^{\mu\nu}u_{\nu}\overline{p}_{i}\;\;\;\;\;\;\;i=1,2,3\label{a_37}
\end{equation}
One can calculate the 4-point function with these variables and it
takes the covariant form
\begin{eqnarray}
i\Gamma^{\mu\nu\lambda\rho (\beta)}(p, q, r) & = & \left(\delta
  (\omega_{1}-\overline{p}_{1})\delta
  (\omega_{2}-\overline{p}_{2})\delta (\omega_{3}-\overline{p}_{3}) +
  \delta (\omega_{1}+\overline{p}_{1})\delta
  (\omega_{2}+\overline{p}_{2})\delta
  (\omega_{3}+\overline{p}_{3})\right)\nonumber\\
  &   &
  \times\overline{u}^{\mu}(p_{1})\overline{u}^{\nu}(p_{2})\overline{u}^{\lambda}(p_{3})\overline{u}^{\rho}(-(p_{1}+p_{2}+p_{3}))\overline{I}_{4}\label{a_38}
\end{eqnarray}
with
\begin{eqnarray}
\overline{I}_{4} & = & {(2ie\pi)^{4}\over 2^{3}}\int
{d\overline{k}\over
  (2\pi)^{2}}\,\left[\epsilon(\overline{k})\epsilon(\overline{k}+\overline{p}_{1})\epsilon(\overline{k}+\overline{p}_{1}+\overline{p}_{2})\epsilon(\overline{k}+\overline{p}_{1}+\overline{p}_{2}+\overline{p}_{3})\right.\nonumber\\
  &   & \times\left\{n(|\overline{k}|) +
    n(|\overline{k}+\overline{p}_{1}|)+\cdots\right.\nonumber\\
  &   & \;\;\;\; -
  2\left(n(|\overline{k}|)n(|\overline{k}+\overline{p}_{1}|) + {\rm
    all\;quadratic\;permutations}\right)\nonumber\\
  &   & \;\;\;\; +
  4\left(n(|\overline{k}|)n(|\overline{k}+\overline{p}_{1}|)n(|\overline{k}+\overline{p}_{1}+\overline{p}_{2}|)
  + {\rm all\;cubic\;permutations}\right)\nonumber\\
  &   & \;\;\;\; -
  \left. 8n(|\overline{k}|)n(|\overline{k}+\overline{p}_{1}|)n(|\overline{k}+\overline{p}_{1}+\overline{p}_{2}|)n(|\overline{k}+\overline{p}_{1}+\overline{p}_{2}+\overline{p}_{3}|)\right\}\nonumber\\
  &   & + \;\;\;\;\;\;\; \left.{\rm all\;permutations\;of\;}
    (\overline{p}_{1}, \overline{p}_{2},
    \overline{p}_{3})\right]\label{a_39}
\end{eqnarray}

It is a little surprising that the 4-point function does not vanish
while the general theorem in section {\bf 2} would imply so. The
reason for the failure of the general theorem is not hard to see. At
finite temperature, there are additional tensor structures available
such as the velocity of the heat bath. Consequently, the expansion for
the vector field in (\ref{a_7}) is no longer the most general at
finite temperature. In fact, one can write (Here $\tilde{\sigma}$ and
$\tilde{\phi}$ are related to $\sigma$ and $\phi$ in a nontrivial way.)
\begin{equation}
A_{\mu} = {1\over e}(\partial_{\mu}\sigma +
\epsilon_{\mu\nu}\partial^{\nu}\phi + u_{\mu}\tilde{\sigma} +
\epsilon_{\mu\nu}u^{\nu}\tilde{\phi})\label{a_40}
\end{equation}
The simplifications noted in section {\bf 2} do not go through in the
presence of these additional terms and hence the general theorem
fails. (There are many ways to see that velocity dependent terms can
arise in the expansion at finite temperature. The simplest, probably,
is to recall that, at finite temperature, there are, in general, more
than one independent transverse projection operators and some of them
depend on the velocity of the heat bath.) Yet, another way of saying
this is to  note that we can
decompose the velocity four-vector along the parallel and
perpendicular directions with respect to the momenta and hence can
rewrite the expansion also in a generalized form of
(\ref{a_7}). However, such a decomposition involves singular inverses
which invalidate the cyclicity property used in (\ref{a_17}) to set the
higher order terms to zero in the general proof.

It is clear, therefore, that unlike at zero temperature, the higher
order terms do not vanish at finite temperature. This is very much
like the behavior of the $0+1$ dimensional theory \cite{{dunne},
{das5}, {das6}} where, at zero temperature, the effective action is
only linear in the external field while, in the presence of a heat
bath, interactions to all orders are generated. Here, however, we
have the simplification that only even amplitudes are
non vanishing. Furthermore, the dependence on the delta functions is a very particular generalization of the $0+1$ dimensional result \cite{{das5}, {das6}} where the n-point function is proportional to $\delta(p_{1})\delta(p_{2})\cdots \delta(p_{n-1})$ and is a simple consequence of gauge invariance alone. From the calculations presented so far (as
well as from the requirement of vanishing divergence and curl), the
structure of the temperature dependent corrections to the higher point
functions is quite clear. The
covariant form of the $2n$-point function can be written as
follows. Let $p_{1}, p_{2},\cdots , p_{2n-1}$ denote the independent
external momenta. Then, with the generalization of the decomposition
given  in (\ref{a_37}), we can write the $2n$-point function as
\begin{eqnarray}
i\Gamma^{\mu_{1}\cdots\mu_{2n}}(p_{1},\cdots ,p_{2n-1}) & = &
\left\{\left(\delta (\omega_{1}-\overline{p}_{1})\cdots \delta
  (\omega_{2n-1}-\overline{p}_{2n-1})\right.\right.\nonumber\\
   &   & \;\;\;\;\left.\left. + \delta
  (\omega_{1}+\overline{p}_{1})\cdots \delta
  (\omega_{2n-1}+\overline{p}_{2n-1})\right)\right\}\nonumber\\
  &  & \times\overline{u}^{\mu_{1}}(p_{1})\cdots
  \overline{u}^{\mu_{2n}}(-(p_{1}+\cdots
  +p_{2n-1}))\overline{I}_{2n}\label{a_41}
\end{eqnarray}
where
\begin{eqnarray}
\overline{I}_{2n} & = & {(2ie\pi)^{2n}\over 2^{2n-1}}\int
{d\overline{k}\over
  (2\pi)^{2}}\,\left[\epsilon(\overline{k})\epsilon(\overline{k}+\overline{p}_{1})\cdots
  \epsilon(\overline{k}+\cdots +\overline{p}_{2n-1})\right.\nonumber\\
  &   & \times\left\{n(|\overline{k}|) + \cdots + n(|\overline{k} +\cdots
 +\overline{p}_{2n-1}|)\right.\nonumber\\
  &   & \;\;\;\; - 2\left(n(|\overline{k}|)n(|\overline{k}
    +\overline{p}_{1}|)+ {\rm
    all\;quadratic\;permutations}\right)\nonumber\\
  &   & \;\;\;\; +
  4\left(n(|\overline{k}|)n(|\overline{k}+\overline{p}_{1}|)n(|\overline{k}+\overline{p}_{1}+\overline{p}_{2}|)+ {\rm all\;  cubic\;permutations}\right)\nonumber\\
  &   & \;\;\;\; + \cdots\nonumber\\
  &   & \;\;\;\; - \left. 2^{2n-1}n(|\overline{k}|)\cdots
    n(|\overline{k}+\cdots 
    +\overline{p}_{2n-1}|)\right\}\nonumber\\
  &   & \left. \;\;\;\; + {\rm all\;permutations\; of\;}
    (\overline{p}_{1}, \cdots 
    , \overline{p}_{2n-1})\right]\label{a_42}
\end{eqnarray}
Thus, collecting all terms, we can write the full effective action at
finite temperature to be
\begin{equation}
\Gamma[A] = \Gamma^{(0)}[A] +
\sum_{n=1}^{\infty}\Gamma_{2n}^{(\beta)}[A]\label{a_43}
\end{equation}
where
\begin{eqnarray}
\Gamma_{2n}^{(\beta)} & = & {1\over 2n!}\int {d\omega_{1}d\overline{p}_{1}\over
  (2\pi)^{2}}\cdots {d\omega_{2n-1}d\overline{p}_{2n-1}\over
  (2\pi)^{2}}\,(\overline{u}\cdot A)(p_{1})\cdots (\overline{u}\cdot
  A)(-(p_{1}+\cdots +p_{2n-1}))\nonumber\\
   &   & \times\overline{I}_{2n}\left(\delta (\omega_{1}-\overline{p}_{1})\cdots \delta
  (\omega_{2n-1}-\overline{p}_{2n-1}) + \delta
  (\omega_{1}+\overline{p}_{1})\cdots \delta
  (\omega_{2n-1}+\overline{p}_{2n-1})\right)\label{a_44}
\end{eqnarray}
and $\Gamma^{(0)}[A]$ is the effective action at zero temperature
given in eq. (\ref{a_12}).

This shows that, at finite temperature, the effective action has
interactions to all even orders unlike the case at zero
temperature. All these additional, temperature dependent terms, of
course, do not change the current conservation as well as the anomaly
of the theory. However, the presence of such terms raises the
interesting possibility that, unlike at zero temperature, there may be
finite temperature effects giving rise to two loop and higher loop
contributions in the fundamental theory. In fact, such contributions
will correspond to diagrams where two (or more) gauge fields are
contracted in the effective action in (\ref{a_43}, \ref{a_44}) (of
course, we are assuming here that the gauge fields are
dynamical). But, a little analysis would show that every such contraction
would involve a factor (remembering the transversality of
$\overline{u}^{\mu}(p)$)
\begin{eqnarray}
\overline{u}^{\mu}(p) D_{\mu\nu}(p) \overline{u}^{\nu}(-p) \delta
(\omega \mp \overline{p}) & = &
-{1\over p^{2}-m^{2}}\overline{u}^{\mu}(p)\overline{u}_{\mu}(-p)
\delta (\omega \mp \overline{p})\nonumber\\
   & = & -{1\over p^{2}-m^{2}}\left({p^{2}\over \overline{p}^{2}}\right)\delta
   (\omega \mp \overline{p}) = 0 \label{a_45}
\end{eqnarray}
Here $D_{\mu\nu}(p)$ is the propagator for the gauge field,
$m^{2}=e^{2}/\pi$ and  we have
used eq. (\ref{a_27}) in the final step. (As a side remark, we would
like to point out that in $1+1$ dimensions\footnote{We thank
  Prof. J. Frenkel for this observation.},
$\overline{u}^{\mu}\overline{u}^{\nu}$ can be expressed in terms of
the usual transverse projection operator as follows.
\begin{equation}
\overline{u}^{\mu}(p)\overline{u}^{\nu}(-p) = - {p^{2}\over
 \overline{p}^{2}}\left(\eta^{\mu\nu} - {p^{\mu}p^{\nu}\over
 p^{2}}\right)
\end{equation}
This observation makes the derivation of the propagator rather
simple.) This is quite interesting,
for it says that, even though there are higher point functions present
in the effective theory, all the radiative corrections in the
fundamental theory are of order one loop. The consequence of this  is
that the temperature dependent, effective theory obtained in
(\ref{a_43}, \ref{a_44}) is purely classical -- it cannot generate any
quantum correction. This is interesting and is indeed quite unusual
and, as is clear from (\ref{a_45}), is a direct consequence of the
specific dependence on delta functions which is also necessary to
maintain the current conservation as well as the anomaly of the
theory. This is indeed yet a new feature that finite temperature field
theories can have. We also note from (\ref{a_45}) that, for $m=0$,
such a contraction will not vanish. Consequently, in perturbation
theory (where the photon does not have a mass), the higher loop
contributions will not vanish individually. They would, in fact, be
highly infrared singular. However, if the perturbation is summed to
all orders, all such contributions would add upto zero as is clear
from (\ref{a_45}).

\section{Soluble Models:}

As is well known \cite{{hagen1}, {das9}}, once the effective
action for the fermion field in an external Abelian gauge background
is known, various soluble models can be directly studied. Therefore,
we will be rather brief in this section.  First, let us recall that
the Schwinger model \cite{schwing} is defined by the Lagrangian
density 
\begin{equation}
{\cal L} = -{1\over 4} F_{\mu\nu}F^{\mu\nu} + {\overline
  {\psi}}\gamma^{\mu}(i\partial_{\mu} - e A_{\mu})\psi.\label{a_46}
\end{equation}
It is clear, therefore, that integrating out the fermions would lead
to an effective action which is the sum of the kinetic term for the
photons and the effective action derived in (\ref{a_43}). Thus, the
effective action, in
addition to containing the mass term for the photon also contains now
the additional interactions whose properties we have already
discussed.

The general model, in $1+1$ dimensions, \cite{{bass}, {das7},
{hagen2}} is described by the Lagrangian density
\begin{equation}
{\cal L} =  -{1\over 4} F_{\mu\nu}F^{\mu\nu} + {\overline
  {\psi}}\gamma^{\mu}(i\partial_{\mu} - e(1+r\gamma_{5}) A_{\mu})\psi.\label{a_47}
\end{equation}
Here $r$ is an arbitrary parameter and this model is known to reduce
to various soluble models under different limits and reductions. We
note that if we define a new gauge field as
\begin{equation}
B^{\mu} = (\eta^{\mu\nu} + r\epsilon^{\mu\nu})A_{\nu}\label{a_48}
\end{equation}
then, using the two dimensional identities in (\ref{a_4}), it is easy
to show that the fermion part of the Lagrangian in (\ref{a_47})
becomes identical to that for the Schwinger model, but in terms of the
$B_{\mu}$ field. We have already
evaluated the effective action for this and so, expressing 
everything back in terms of the $A_{\mu}$ field, we would have the
effective action for the general model which then, would give the
effective action for various soluble models at finite temperature
under different limits and reductions \cite{old}. We would simply like to note
here that for the gradient coupling model,
\[
A_{\mu} = \partial_{\mu}\phi
\]
namely, the gauge field can be identified with the gradient of a
scalar. In such a case, however, it is clear that
\begin{equation}
\overline{u}(p)\cdot A(p) = 0\label{a_49}
\end{equation}
Consequently, all the temperature dependent corrections to this model
identically vanish and  the zero temperature effective action is the
full action (independent of the regularization used).

\section{Conclusion}

In this paper, we have studied, systematically, the effective action
for $1+1$ dimensional, massless fermions interacting with an external Abelian
gauge field at finite temperature. While the naive expectation would
be that only the two point function is corrected by temperature, we
have calculated and shown that the effective action, in fact, contains
interaction terms to all (even) orders. The exact form of the
effective action is obtained and it is shown that these additional
temperature dependent terms do not change the anomaly or the current
conservation. The non-analytic structure of the effective action at
finite temperature is pointed out. It is also pointed out that these
temperature dependent terms in the effective action have a very
specific structure which prevents them from generating any quantum
mechanical correction. To the best of our knowledge, this is  a
new feature of field theories at finite temperature. The solubility of
various two dimensional models is also briefly discussed. We hope that
some of the features found here will help in the understanding of the
structure of the effective action for a fermion interacting with an
arbitrary gauge field in $2+1$ dimensions. As a final comment, we would like to add that we have also calculated the physical, retarded Greens function \cite{das1} in this model. All the temperature dependent parts vanish which is consistent with our conclusion that these new terms in the finite temperature effective action cannot lead to any quantum correction \cite{new}.

We would like to thank Profs. J. Frenkel and M. Gomes for many
discussions. A.D. would like to thank the members of the Mathematical
Physics Department  of 
USP for hospitality where this work was done. A.D. is supported in part by US 
DOE Grant number DE-FG-02-91ER40685, NSF-INT-9602559 and  FAPESP. 
A.J.S. is partially supported by CNPq (the National Research Council of 
Brazil).

\end{document}